\documentclass[a4paper]{conm-p-l}

\usepackage{setspace}
\usepackage{mathrsfs}
\usepackage{amssymb}
\usepackage{amsmath}

\usepackage[all,cmtip,knot]{xy}
\entrymodifiers={+!!<0pt,\fontdimen22\textfont2>}

\usepackage{tikz}

\def\beq{\begin{equation}}
\def\eeq{\end{equation}}

\def\Cinf{C^\infty}
\def\R{\mathbb{R}}
\def\C{\mathbb{C}}
\def\Z{\mathbb{Z}}

\def\Span{\qopname \relax o{span}}

\def\goesto{\rightarrow}
\def\isomto
{\mathrel{\tilde{\rightarrow}}}

\DeclareMathOperator{\End}{End}
\DeclareMathOperator{\Aut}{Aut}

\def\GL(#1,#2){{\mathrm{GL}_{#1}#2}}
\def\SL(#1,#2){{\mathrm{SL}_{#1}#2}}
\def\Sp(#1){\mathrm{Sp}(#1)}
\def\sp(#1){\mathfrak{sp}(#1)}	

\usepackage{graphicx,graphics}
\usetikzlibrary{intersections}

\def\bdy{\partial}

\usepackage{fullpage}
\usepackage{stackengine}
\usepackage{url}

\renewcommand{\emph}{\textsl}
\newcommand{\vect}{\mathbf}

\def\Hilb{\mathscr{H}}
\def\PS{\mathscr{S}}
\def\Proj{
{P}}
\def\vac{\ket{\emptyset}}

\def\arquant{\raisebox{-2pt}{\begin{tikzpicture}
\draw[->,double] (0,-0.4) -- (0.4,-0.4);
\end{tikzpicture}}}
\def\artrunc{\raisebox
{0pt}{\begin{tikzpicture}
\draw[->,dashed] (0,-0.4) -- (0.4,-0.4);
\end{tikzpicture}}}
\def\arcoup{\raisebox
{0pt}{\begin{tikzpicture}
\draw[->] (0,-0.4) -- (0.4,-0.4);
\end{tikzpicture}}}
\def\arcont{\raisebox
{0pt}{\begin{tikzpicture}
\draw[->,dotted] (0,-0.4) -- (0.4,-0.4);
\end{tikzpicture}}}
\usepackage[italicdiff]{physics}
\usepackage{tikz,tikz-cd}

\title{An introduction to spin systems 
for mathematicians}

\author{Ingmar Saberi}
\address{Mathematisches Institut, Ruprecht-Karls-Universit{\"a}t Heidelberg,
69120 Heidelberg, Germany}
\email{saberi@mathi.uni-heidelberg.de}

\date{First begun: August 13, 2017. Last compiled: \today.}

\begin{document}

\maketitle

\nocite{*}

\section*{Introduction}

It seems to have become a fashionable truism in some mathematical circles that physics is impossible to understand. This is despite a deep and growing relationship between the two fields;
physical reasoning has come to be viewed as some sort of mysterious oracle, which from time to time emits an interesting and profound idea such as mirror symmetry, but whose inner workings are beyond comprehension.

This is, of course, not true; physicists, after all, are made rather than born, and the theorists among them learn their trade in a way that is (to any of the enormous number of people without interest in either discipline) indistinguishable from a mathematician's training. 
The real stumbling block in establishing mutual understanding between these two camps seems to lie in the radically different ways we are taught to position ourselves in an epistemological relationship to the world. For a mathematician, every system is closed: one does not care in what kind of system one is working, as long as the axioms are clear and things are well-defined. Once the assumptions are fixed, the borders of the system cannot change.

For a physicist, on the other hand, every system is open, and (more to the point) approximate. One never really expects that the mathematical problem one formulates and then solves will provide an \emph{exact} or \emph{complete} description of a physical system. Rather, it is some kind of picture of a real situation, which represents some features and perhaps leaves some others out. Many such pictures of the same situation may exist; one expects a relationship between them because they are pictures of the same situation, rather than because of anything intrinsic to either picture in itself. 

Speaking narrowly, the purpose of this contribution is to give an introductory discussion of spin systems, emphasizing why such systems are of physical interest and what sorts of questions about them physicists would like to have answers to. This 
motivates some discussion of the notion of a \emph{phase}, and so serves to introduce the central theme of the conference, 
Dan Freed's lectures on recent work using ideas from homotopy theory and topological field theory to compute groups of invertible phases. More broadly, though, we would like to try and cast at least a bit of light on how spin systems and field theories can (and often do)  appear interchangeably. The central reason for this is that they are two different pictures of the same situation.

One way of expressing what that situation is would be to say that physicists care about field theory because it is a way of talking about systems with \emph{many} degrees of freedom, labeled by points in a geometric space, and interacting in a way that respects the notions of homogeneity and locality associated to the geometry of that space. Examples of such systems, like Maxwell's electromagnetism, are all around us, but spin systems are also models of many degrees of freedom that are designed to respect the same notions.  This idea is perhaps unfamiliar to those whose road into field theory starts with TQFT (in which there are no local degrees of freedom present at all), but it underlies the way physicists switch back and forth between various classes of models, secure in the belief that they represent similar physics. This constant change of perspectives can be a bit nauseating for those who are not used to it, and so we try to offer some discussion as to  why one might believe it's reasonable. In a style typical of physics, this discussion will proceed through a series of examples. 

We've also tried to highlight the importance of symmetry to the study of phases, and show why it is so crucial that one's concept of symmetry be expanded to include algebras of operators that are nonlocal. We feel that this is the most straightforward way to understand the crucial role of anyon systems in $2+1$-dimensional physics, as well as gesture to the importance of line and surface operators in the study of field theory in general. As such, we have tried to emphasize the role of anyons, and show their relationship to algebras of nonlocal symmetries.




Essentially none of the material in this article is new. The particular choice of topics and their presentation, however, is novel, and we feel it fills a gap in the existing literature. Most existing treatments of the material we discuss are either too elementary or too dependent on physical language and intuition to be of great use to a mathematically literate audience who are nonetheless physical novitiates. With this in mind, we have chosen to include a brief exegesis of topics (including basic quantum mechanics)  that are found in countless standard physics textbooks. Emphasis has been placed on key ideas (``system, state, measurement'') and pieces of physical or epistemological reasoning that have precise meaning, but belong a bit more to the realm of natural philosophy than to mathematics \emph{per~se}.

A couple of points on notation: we will follow conventions that are more typical of the physics literature than of mathematics, in the hope that this will prove to be instructive exposure rather than merely a source of confusion.  As such, letters in bold type will denote spatial vectors. Indices from the middle of the Roman alphabet ($i$, $j$, and so on) will denote the components of such a vector, and so can be thought of as belonging to the set $\{1,2,\ldots,d\}$, where $d$ is the spatial dimension of the system at hand; we will also sometimes denote the index set by $\{x,y,z\}$.  $i$ will therefore appear both as an index and as the imaginary unit; it should always be clear from context which use is intended. Indices from the middle of the Greek alphabet ($\mu$, $\nu$, \ldots) will denote the components of ``four-vectors,'' i.e., sections of the tangent bundle to the spacetime. One will often use the same letter in both contexts, for both a four-vector and its spatial part; the index value zero refers to time, so that expressions like 
\[
x^\mu = (x^0,\vect{x})
\]
are common. Greek indices from the beginning of the alphabet ($\alpha$, $\beta$) will label lattice sites; we will reserve Latin indices from the beginning of the alphabet ($a$, $b$, \ldots), as well as other Greek letters such as~$\gamma$, for other index sets, to be defined specific to a given context.

Index placement is used to distinguish vectors from dual vectors; thus, an object like $x^\mu$ is associated to the tangent bundle, and $p_\mu$ to the cotangent bundle. The evaluation map is denoted by an implicit summation, of the form $p_\mu x^\mu$, which is to be understood whenever repeated indices appear. Often, indices are ``raised and lowered'' using a metric, whose components are denoted $g_{\mu\nu}$; thus, for example, the expression $x_\mu$ is understood to mean $g_{\mu\nu} x^\nu$. Again, both raised indices and exponents will appear without further comment, but context should be a sufficient guide. We will carefully obey the index placement rule for four-vectors, but will be sloppy about it otherwise (for example, with quantities like the angular momentum, which are not, strictly speaking, spatial vectors, although often represented as such).

A similar convention differentiates between vectors and dual vectors in a quantum-mechanical context: we frequently write elements of the Hilbert space~$\Hilb$ with symbols of the form $\ket{a}$, and denote elements of~$\Hilb^\vee$ by~$\bra{b}$. The evaluation pairing then looks like $\bra{b}\ket{a}$, while a symbol of the form $\dyad{a}{b}$ denotes the obvious element of~$\End(\Hilb) \cong \Hilb \otimes \Hilb^\vee$. When not clear from context, ``operators''---elements of~$\End(\Hilb)$---may wear a circumflex. The labels we have written here as $a,b$ may be any labels or indices characterizing the intended vector, but will often be eigenvalues of a particular operator labeling its eigenvectors. Again, letters may be used in more than one of these contexts simultaneously, so that an eigenvalue equation may appear something like
\[
 \hat{a} \ket{a} = a \ket{a}. 
 \]
A similar convention extends to the action of operators on arbitrary states, so that one also sees equations like 
\[
\hat{a}\ket{\psi} = \ket{\hat{a}\psi}
\]
(which should be understood as defining the meaning of the expression at right).

These conventions, or slight variants, are very commonly (though not universally) obeyed, and are well-known enough as to be used in many papers without explanation. For that reason, we hope that the brief leitfaden and sample usage in this article will render further encounters with the physics literature less alienating.

\section{General principles of quantum mechanics: a reminder}
\label{sec:QM}

We begin our discussion with a small crash course in the principles and axioms of quantum mechanics. In keeping with the idea that the relation between quantum and classical physics is crucial for understanding the former, we include some discussion of classical systems, hoping to emphasize physical ways of thinking that may be  less familiar to our intended audience, as well as the epistemological concepts and ideas or intuitions about nature that motivate the use of various mathematical structures.

The concept of \emph{state} in physics connotes a set of information about a system, which is sufficient to completely determine the future time-evolution of that system, at least to the greatest extent possible within the theory being considered. The theory then consists of a definition of the state of a system, together with the set of rules that govern its time-evolution. Since measurement is the only possible epistemic interaction one can have with the system, the state is something that can be determined in principle by repeated measurements, in any fully satisfactory theory of physical phenomena. 
In quantum mechanics, the requirement continues to hold, up to an important subtlety about how one should interpret the word ``determined;'' we will return to this subtlety shortly.

The space of states, therefore, admits a one-parameter group of automorphisms, representing evolution in time. (If information is lost under time-evolution, so that only a semigroup of endomorphisms corresponding to \emph{forward} time-evolution is well-defined, one might say that the system ``lacks \emph{unitarity;}'' we will ignore such systems in what follows.) Points in the state space (at a particular moment in time) are therefore in one-to-one correspondence with trajectories of the time evolution.

Two other concepts that are important are  \emph{reductionism} and \emph{locality}. Reductionism means that it should be possible to separate any system into constituent pieces or subsystems, such that the total dynamics can be described in terms of the independent dynamics of the subsystems, together with interactions between them. Locality is the principle that the interactions between subsystems should become negligible when the subsystems are sufficiently far apart, so that it is possible to model a physics experiment taking place in a room without writing the dynamics of other nearby buildings or of the Andromeda galaxy: the subsystem can be studied on its own. As such, these notions are critical for any workable and predictive theory: they ensure that we do not need all information about the whole universe in order to make predictions about some small part of it. So we should have a notion of forming a composite system out of two subsystems, and a notion of a total dynamics in which the subsystems do not interact.

To give an example, let's review how these notions are manifested in classical physics. The reader is referred to the literature for more detail, and in particular to~\cite{arnold, landau}.

The classical state of (for example) a point particle consists of its position and velocity, and the way in which these are measured is taken to be obvious. The rule for the evolution of the state is Newton's equation, $
{F} = m
{\ddot{q}}$. Knowledge of the position and velocity is sufficient initial data to specify, by solving Newton's equation, the position $
{q}(t)$ at all later times. Of course, one then also knows  the velocity $
{\dot{q}}(t)$; in other words, the evolution of the state in time is completely determined. 

There are numerous ways of reformulating classical mechanics that demonstrate more closely its connections to quantum mechanics; while these reformulations are equivalent in principle, they highlight different important features of and perspectives on the theory. One of these is the \emph{principle of least action,} which states that the time-evolution takes place along a trajectory that extremizes the \emph{action functional}
\beq
 S = \int dt\, L(
{q},\dot{
{q}}).
 \eeq
Here the function $L$ is called the \textsl{Lagrangian}. For a particle moving in a potential, it is the difference of the particle's kinetic and potential energies: 
\beq
 L = \frac{1}{2} m \dot{
{q}}^2 - U(
{q}),
\eeq
where $U$ denotes potential energy. It is an instructive exercise to check that the condition for~$S$ to be extremal (the Euler-Lagrange equation) is then nothing other than Newton's equation for the conservative force $F = - \pdv*{U}{q}$.

Another formulation, which is especially instructive for our purposes, requires taking the Legendre transform of the Lagrangian to obtain another functional called the \emph{Hamiltonian}. 
That is, one defines a new coordinate on the state space by the relation
\beq
p = \pdv{L}{\dot q},	\label{def-p}
\eeq
and then sets 
\beq
H(q,p) = p\dot{q} - L(q,\dot{q}), \label{def-H}
\eeq
where it is implicit that $\dot{q}$ on the right-hand side is eliminated in favor of~$p$, using~\eqref{def-p}. After doing this, the second-order equations of motion appearing above are replaced by a set of two first-order equations:
\beq
 \dot{q} = \pdv{H}{p}, \quad  \dot{p} = - \pdv{H}{q}.
\eeq
The first of these is immediate from~\eqref{def-H}, and merely re-expresses the constraint~\eqref{def-p} defining the canonical momentum $p$. The second is Newton's equation, rewritten in the first-order form $F = \dv*{p}{t}$. 

Hamilton's formulation has several advantages. First off, it shows that the space of states, or phase space~$\PS$, is symplectic: $q$ and $p$ are Darboux coordinates, and the time evolution is Hamiltonian flow. Secondly, it gives a clear notion of observables, which are simply functions on the phase space. The space of observables therefore carries a Poisson bracket, which in Darboux coordinates is simply expressed as 
\beq
\{f,g\} = \qty( \pdv{f}{q} \pdv{g}{p} - \pdv{f}{p} \pdv{g}{q} ).
\label{def-PB}
\eeq
(In cases where more than one pair of variables is present, a sum is understood.) Here $f$ and $g$ are any elements of the space of observables, $\Cinf(\PS)$. It is important to note that~$\{q,p\}=1$.
What is more, the dynamics of an arbitrary observable are simple to state in this language: they are easily written as
\beq
\dv{f}{t} = \{f,H\}.
\label{eq:Ham-eq}
\eeq
Again, it is an instructive exercise to work out these statements in detail.
Constants of the motion, or conserved quantities (which are by definition observables whose value does not change under time evolution) are therefore observables that Poisson-commute with the Hamiltonian. A system is said to be \emph{completely integrable} when $n$ mutually commuting such quantities can be found ($2n$ being the dimension of the phase space).

Furthermore, it is clear what it means to concatenate two subsystems in a noninteracting fashion: since the state is characterized by a pair of states, one for each subsystem, the phase space is the Cartesian product, and the noninteracting dynamics are governed by the Hamiltonian $H_1+ H_2$.

We can now turn to how the same epistemological ideas are represented in a quantum-mechanical theory. Luckily, much of the Hamiltonian formalism we've reviewed can be carried over to quantum mechanics in a way that makes the parallelism or correspondence clear. We'll start at the beginning,  with a reminder of how the concept of the state is represented in quantum theories. It is helpful to introduce the notion of state and the notion of measurement (which, of course, are closely related) side by side.

Let us suppose that some quantity $a$ can be measured, and the result of this measurement is an element of a set $\Sigma=\{a_\gamma\}$ of possible outcomes. ($\Sigma$ is a subset of~$\R$, which may be continuous or discrete; for clarity, we will treat the discrete case, although the principles remain the same.) Suppose, furthermore, that a measurement of~$a$ completely characterizes the system: there are no other quantities available to measure. Such a system might be, for example, an electron, where what is measured is its spin along a fixed axis; $\Sigma$ then consists of only two values, ``up'' and ``down'' (or $\pm 1$).\footnote{It should be carefully noted that, in cases where more than one measurement must be made, $\Sigma$ can consist of a \emph{smaller} set of measurements that those that define $\PS$; the source of this subtlety should become clear momentarily.}

The source of essentially all of the strangeness in quantum mechanics is the following postulate:
Given a complete set $\Sigma$ of possible outcomes for measurements, the quantum-mechanical state space is not $\Sigma$ itself; rather, it is the formal linear span of $\Sigma$ over~$\mathbb{C}$, subject to the constraint of $L^2$-normalizability when $\Sigma$ is infinite. To speak more properly, it is the space $\Hilb=L^2(\Sigma)$ of normalizable complex-valued functions on that set. This space is equipped with an orthogonal decomposition as a direct sum of one-dimensional subspaces, spanned by the elements of~$\Sigma$, and labeled by the corresponding real numbers $a_\gamma$ which are the outcomes of the measurement. A decomposition of a complex vector space into an orthogonal direct sum labeled by real numbers is equivalent to the data of a Hermitian operator acting in that space; we are thus led to the idea that observables in quantum mechanics are represented by Hermitian operators. 

This is, on the one hand, a familiar kind of statement to mathematical ears. It is a very standard gambit to linearize a problem by thinking about a space of functions: the functions on, for example, a set or topological space form a vector space, often equipped with additional algebraic structure, and a group action on the set gives a linear $G$-representation on the function space. So one might be tempted to think of the passage to quantum mechanics as a way of replacing a symplectic geometry problem by a (potentially easier) linear algebra problem, and this is in some loose sense true.

On the other hand, this postulate is an overwhelmingly peculiar and counterintuitive thing to say about the physical world. For instance, it immediately follows from this that even a system which is completely determined by a single measurement admitting only two possible outcomes has \emph{uncountably} many distinct possible states. 

The attentive reader will immediately point out a problem here. We've said that the state is determined in principle by measurements; how can a system in which any measurement has at most two outcomes have more than two possible states? 

The key lies in the identification of measurable quantities with Hermitian operators, or, what is the same, with a choice of basis in the Hilbert space and a marking thereof by outcomes of the measurement. While the dimension of the Hilbert space indicates the number of possible outcomes for any given measurement, different choices of basis are always possible and correspond to physically distinct measurements. Furthermore, these different measurements are \emph{mutually incompatible:} being in an eigenstate of one such operator implies that one is not in an eigenstate of the other, and therefore that the other takes on no particular definite value.

Of course, a measurement can still be made in such a state; how is one to understand what its outcome will be? The answer is provided by perhaps the most surprising postulate of quantum mechanics, the \emph{Born rule}. Its content is as follows: Suppose an observable $\hat{a}$ is to be measured, with possible eigenvalues $a_\gamma$. Imagine that a system is prepared in the state 
\[
\ket{\psi} = \sum_\gamma c_\gamma \ket{a_\gamma},
\]
for some constants $c_\gamma$. Following this, $\hat{a}$ is measured. The outcome will be one of the $a_\gamma$; if the measurement is repeated, the probability of obtaining $a_\gamma$ will be $|c_\gamma|^2$. 
Furthermore, \emph{after} the measurement is performed, the state of the system is $\ket{a_\gamma}$, as evidenced by the measured value of~$\hat{a}$. 

Thus, measuring a state in quantum mechanics doesn't tell you what the state was previously; rather, it \emph{prepares} the state corresponding to the outcome of the measurement. Nonetheless, a general state~$\psi$ can be determined, assuming many identical copies of it can be obtained or similarly prepared: repeated measurements obviously identify the values of~$\abs{c_\gamma}$, while other experiments are possible in principle to access the relative phase of~$c_\gamma$ and~$c_{\gamma'}$. For the sake of space, this is all we can say about this subject here.

As a simple example of this framework, if the Hilbert space is two-dimensional, any two-by-two Hermitian matrix is an observable. The space of such matrices is spanned 
(along with the identity, which is a trivial measurement) by the \emph{Pauli matrices:} these are a set of standard Hermitian matrices $\sigma_i$, obeying the commutation relations
\[
[\sigma_x, \sigma_y] = i \sigma_z
\]
(together with all cyclic rearrangements). The $i\sigma_i$ thus form a basis of~$\mathfrak{su}(2)$.
By convention, they are written in an eigenbasis of~$\sigma_z$, so that 
\[
\sigma_x = \frac{1}{2} \begin{bmatrix} 0 & 1 \\ 1 & 0 \end{bmatrix}, \quad
\sigma_y = \frac{1}{2} \begin{bmatrix} 0 & \,\,\llap{$-$}i \\ i & \,\, 0 \end{bmatrix}, \quad
\sigma_z = \frac{1}{2} \begin{bmatrix}1 & \,\, 0 \\ 0 & \,\,\llap{$-$}1 \end{bmatrix}.
\]
Reader beware: our conventions for Pauli matrices differ by a normalization factor of $1/2$ from those most typical in the literature!
The algebra of all observables in such a system is thus equivalent to~$\mathfrak{su}(2)$, which is sometimes called the \emph{angular momentum algebra}, and the Hilbert space is its defining representation, often termed the \emph{spin-$1/2$ representation} in physics.
One thus often thinks of such a two-dimensional Hilbert space as representing the spin of an electron, or any other spin-$1/2$ particle in three-dimensional space. The conclusion is that the spin of such a particle may be measured along any axis in three-dimensional space---but with the understanding that any two such measurements along non-collinear axes are \emph{mutually incompatible}, and states of definite (for example) $x$-spin do not correspond to states with any definite $z$-spin. For further discussion of the physical behavior this implies, and the experiments that confirmed it, we recommend the discussion of the Stern--Gerlach experiment in~\cite{sakurai}. The existence of non-basis states implies even stranger phenomena when a system can be decomposed into subsystems; these go by the name of \emph{entanglement}, and we will return to them shortly.


What does the  algebra of the Pauli matrices have to do with  angular momentum? Well, it's obviously the Lie algebra of the three-dimensional rotation group, but the same algebra can also be derived by remembering that angular momentum in three dimensions is defined by the equation
\[ \vect{L} = \vect{q} \times \vect{p},\]
and computing the Poisson brackets of the components of this pseudovector using~\eqref{def-PB}. One finds that 
\beq
\{ L_x, L_y \} = L_z.
\eeq
So the failure of the components of angular momentum to commute was \emph{already present} in classical mechanics, reflecting the fact that rotations around different axes don't commute, but encoded in subtle fashion in the Poisson bracket. Even the famous canonical commutation relation, implying the uncertainty relation between position and momentum, was already written down above, in its classical form $\{q,p\} = 1$!
Quantization thus asks (among other things) for a deformation of the algebra of classical observables, which was just~$\Cinf(\PS)$, to a noncommutative algebra with commutation relations defined by the rule
\beq
[\hat f, \hat g ] = i \hbar \{ f, g\} 
\eeq
(at least at leading order in~$\hbar$). 

Even the rules of time evolution can be obtained by this method: from~\eqref{eq:Ham-eq}, we obtain
\beq
i \hbar \dv{\hat f}{t}  = [\hat f, H],
\label{eq:SE}
\eeq
which is equivalent to the Schr\"odinger equation. One usually thinks of time-evolution as acting on the states, but of course it is equally natural to think of a one-parameter subgroup of~$\Aut(\Hilb)$ acting on~$\Hilb$ directly, or by conjugation on~$\End(\Hilb)$---i.e., on operators instead of states. Since one can only measure things of the form~$\bra{\psi}\hat{a}\ket{\phi}$, the difference is immaterial, like that between active and passive coordinate transformations; in physics, one refers to these two formulations as the \emph{Schr\"odinger} and~\emph{Heisenberg pictures}. And, of course, the content of~\eqref{eq:SE} is the same in both pictures; it merely says that $H$ is the generator of time-evolution, i.e.\ that $i\hbar \pdv*{}{t} = H$, where the action of~$H$ is interpreted either  in the fundamental representation or in the adjoint action on~$\End(\Hilb)$. 

A couple of further subtleties need to be remarked on immediately. First of all, a state is not an element of~$\Hilb$; rather, it is a one-dimensional subspace. So the space of physical states is really the space of lines in~$\Hilb$, i.e., its projective space $\Proj\Hilb$. But a theorem of Wigner ensures that symmetries of~$\Proj\Hilb$ can always be lifted to~$\Hilb$, so that they are  linear and unitary (or possibly  antilinear and antiunitary). So it is typical to always think in terms of~$\Hilb$, while remembering that scalar multiplication is an equivalence relation on physical states.

Secondly, we assumed for the sake of exposition that an allowed value $a_\gamma$ of the measurement we called~$\hat a$ characterizes one and only one state. But we may actually need to make more than one distinct measurement in order to specify the state fully. One typically assumes that there exists a \emph{complete set of commuting observables}, $\hat{a}^{(k)}$, such that one can label every state uniquely by a list of their simultaneous eigenvalues: i.e., 
\beq
\Hilb =L^2(\Sigma) =  \bigoplus_{k,\gamma} \C \ket{a_\gamma^{(k)}}. 
\eeq
Of course, the operators $\hat{a}^{(k)}$ can be simultaneously diagonalized (as is done above) if and only if they mutually commute. The existence of the CSCO is the postulate that there exists a set of compatible measurements which serve to fully characterize the state space.
The Born rule also needs slight generalization: if the eigenspace corresponding to~$a_\gamma$ is degenerate, the probability will be $\bra{\psi}P_\gamma\ket{\psi}$, where $P_\gamma$ is the projection operator onto the eigenspace. Since $\sum_\gamma P_\gamma = 
{1}$, it is immediate that the probabilities of all possible measurements sum to unity.

Thirdly, one needs to be careful: $\Hilb$ is emphatically \emph{not} the space of $L^2$ functions over the phase space! The issue, of course, is that pairs of coordinates like $q$ and~$p$ have a canonical nonzero Poisson bracket, and so fail to commute; one can choose only \emph{one} of such a pair as part of a set of commuting observables. This is, of course, related to the Stone--von~Neumann theorem, which says that the algebra of a canonical pair of real variables has a unique unitary representation on~$L^2(\R)$. The consistent choice of ``half'' of the phase space variables, in order to construct the Hilbert space, is a source of difficulty in quantization of general symplectic manifolds; the additional data required is sometimes called a~\emph{polarization}.

Finally, let us remark on how one forms composite systems in quantum mechanics. Armed with the ideas we have developed, it is easy to see how this should go: If measuring $\hat{a}$ completely characterizes one subsystem, and measuring $\hat{b}$ completely characterizes the other, then a measurement of both of these should characterize their composite. We expect that the two measurements are compatible because each pertains only to one of the subsystems, and it is supposed that the two could in principle be isolated from one another. The joint Hilbert space is therefore the $L^2$-span of the $\ket{a_\gamma,b_\zeta}$---but this is just the tensor product of~$\Hilb_1 = \Span \ket{a_\gamma}$ and~$\Hilb_2 = \Span \ket{b_\zeta}$. (We are passing over issues arising in infinite dimensions in silence; our aim is to give an intuitive idea as to \emph{why} the tensor product is the correct structure to use.) 

Measurements on one subsystem that do not affect the other should then be represented by the relevant operator on the subsystem, tensored with the trivial operator on the other; this means that it will act correctly on the basis vectors of a tensor product basis, returning the value of the measurement independent of the state of the other system. The natural dynamics for a noninteracting pair, then, is just the sum of the two Hamiltonians, where each is taken to act in this way on the entire system:
\beq
H = H_1 \otimes 1 + 1 \otimes H_2.
\label{eq:sumH}
\eeq 
(We will often write this and similar expressions in the abbreviated form $H = H_1 + H_2$ in what follows, the natural maps between spaces of operators being understood implicitly.)

Of course, just as the Hilbert space contained many states that were not eigenstates for an operator, and so (with strange and counterintuitive consequences)  didn't correspond to any particular value of the corresponding measurement, the composite Hilbert space $\Hilb = \Hilb_1 \otimes \Hilb_2$ contains many states that are not basis states in the tensor product basis. Moreover, many of these states can't be basis states in \emph{any} tensor product basis---just as rank-one operators span, but do not exhaust, $\End(\Hilb) = \Hilb \otimes \Hilb^\vee$. Physicists refer to such states as \emph{entangled} between the two subsystems, and entanglement remains an inexhaustible source of strange behavior and theoretical interest even in the present day. We will not be able to say much about it here, but we feel it is important to note in closing that entanglement is connected to interactions between subsystems. If the Hamiltonian takes the non-interacting form~\eqref{eq:sumH}, it is obvious that it can be diagonalized in a tensor-product basis, between the energy eigenstates of each subsystem. Thus, all eigenstates of the composite Hamiltonian have zero entanglement. Conversely, when energy eigenstates are entangled, it follows that there are interactions between the subsystems. There has been great interest in developing quantitative measures of entanglement, and exploring their properties in energy eigenstates of interacting systems, such as the vacuum state of a quantum field theory (see Chapter 5 of~\cite{WenBookNew} and references therein).

\section{A few words about field theories}
\label{sec:FT}

At root, a \emph{field theory} is a theory whose dynamical degrees of freedom are fields, and \emph{fields} are quantities (like the electric field, the air pressure, or the temperature) that can be measured independently throughout space. It is common to make further requirements, for example that the field theory be \emph{Lorentz-invariant}, i.e.\ compatible with the principle of special relativity. But requirements like these are not inherent in the concept, and there are interesting examples where they are not met. Other than the fact that the degrees of freedom are labeled or indexed by points in space, all of the generalities of \S\ref{sec:QM} are equally true for field theories.

We will make one further assumption, though, which we already identified in the very beginning of the article: namely, that the theory exhibits \emph{locality}. In other words, the action functional that encodes the theory's dynamics should be able to be computed in a local fashion. For a typical continuum field theory, this will amount to the statement that
\[
S = \int dt\, L = \int dt\, d\vect{x}\, \ell[q,\dot{q}],
\]
where $\ell$, the \emph{Lagrangian density}, depends on the fields $q(\vect{x})$ and their derivatives at the same point of space only.
Similarly, field theories (just like any other theories) admit a Hamiltonian operator, and locality means that it is likewise the integral over space of a Hamiltonian density:
\[
H = \int d\vect{x}\, h[q, p].
\]
For discretized field theories and spin systems, the concept of locality will be implemented in a slightly different fashion, but its essential meaning is the same.

A further assumption we can make is that $\ell$ and~$h$ depend only on the values at a given point, and not explicitly on the point (the value of~$\vect{x}$) itself. This is an assumption of \emph{translation invariance} or \emph{homogeneity:} it means that the dynamics at each point in space are the same.

The most basic example satisfying these requirements is a free scalar field theory: its degree of freedom is a single real-valued degree of freedom $q(\vect{x})$, whose dynamics are determined by the functional
\beq
S = \frac{1}{2} \int dt\,d\vect{x}\, \left( \partial^\mu q \partial_\mu q + m^2 q^2 \right)
\eeq
After an integration by parts, we can write it as 
\beq
S = \frac{1}{2} \int dt\,d\vect{x}\, q \left( - \partial^\mu \partial_\mu + m^2 \right) q,
\eeq
and it is a simple exercise to check that the associated equation of motion is the \emph{Klein--Gordon equation:}
\beq
\qty(\pdv[2]{}{t} - \laplacian + m^2)q(t,\vect{x}) = 0.
\eeq
We'll look at this equation in a little more detail, in the simplest possible case: namely, in one spatial dimension, setting the parameter $m^2=0$. Then it reduces to the wave equation,
\beq
\qty( \pdv[2]{}{t} - \pdv[2]{}{x} ) q(t,x) = 0.
\label{eq:wave}
\eeq
Of course, solving this equation is a triviality: because it's translation-invariant and time-independent, we can also diagonalize translations. This amounts to asking that $q(t,x)$ is also an eigenfunction of the translation operators $\dv*{}{x}$ and~$\dv*{}{t}$, in other words, a plane wave:
\beq
q(t,x) \propto \exp i (kx-\omega t).
\label{eq:ansatz}
\eeq
The wave equation then becomes an algebraic relation (the \emph{dispersion relation}) on the eigenvalues of these operators: it reads simply $k^2 = \omega^2$, which is linear and expresses the fact that waves can move in either direction at a speed independent of their frequency. (Here we've tacitly set this speed to $1$. The fact that the speed of light is constant reflects the fact that, in the absence of matter, electromagnetism is a free theory very similar to this one.) The essential meaning of the word ``free''\footnote{There is an unfortunate clash of vocabulary here. ``Free'' objects, in mathematics, tend to be the most general possible examples of their kind, so that all others arise as quotients, or (what is the same) by introducing additional relations. In physics, a ``free'' theory is one without interactions, i.e., one where the scattering is in some sense the \emph{least} general possible. The author is grateful to B.~Knudsen for related discussion that drove this point home.} is that the equation is linear: the superposition of any two solutions is again a solution, which means that waves can pass through one another and neither is affected by the other's presence. No scattering, or interaction, of any kind takes place. 

Let's now imagine discretizing this system: instead of the degree of freedom $q$ being defined for all values of~$x$, we'll imagine that it can be measured only for~$x=\alpha a$, where $\alpha$ is an integer labeling lattice sites, and $a$ is a new dimensionful parameter that measures the lattice spacing. An immediate analogue of~\eqref{eq:wave} is obtained by replacing spatial derivatives by difference operators:
\beq
\pdv[2]{}{t} q_\alpha(t) + \frac{1}{a^2}\left(2q_\alpha(t) -  q_{\alpha+1}(t) -  q_{\alpha - 1}(t)\right) = 0.
\eeq
This equation is still local, in the sense that a particle only interacts with its nearest neighbors. And it is solved by the same ansatz as~\eqref{eq:ansatz}: we diagonalize time translations and (discrete!) lattice translations, choosing eigenfunctions 
\beq
q_\alpha(t) \propto \exp i(k a \alpha - \omega t).
\eeq
It is then trivial to see that the new dispersion relation is 
\beq
\omega^2 = \frac{2}{a^2} \qty(1 - \cos(k aL) ) = \frac{4}{a^2} \sin[2](k a/2).
\eeq
As the reader will have no trouble checking, this reduces precisely to the linear dispersion we found before in the limit $k a \ll 1$; however, it deviates from it when $k \sim 1/a$, i.e., at sufficiently short wavelengths. And this makes sense: short-wavelength waves will be not be represented well by discrete sampling, just as high frequencies on a CD are distorted due to digitization noise. Taking the continuum limit, $a \goesto 0$, recovers the continuous wave equation we studied previously.

Another important effect is that the range of allowable wavelengths is no longer infinite: the parameter $k$ is now circle-valued, expressing the fact that the Pontryagin dual of~$\Z$ is the circle group, and this has the effect of imposing a hard momentum cutoff at~$k = \pi/a$. This is often called an \emph{ultraviolet cutoff} (UV) in physics, and eliminates one common source of divergences in field theories. However, some infinities may remain, because we are still dealing with a system with infinitely many degrees of freedom: the parameter $k \in [-\pi/a,\pi/a]$ labeling independent solutions is continuous. To have a system with only finitely many degrees of freedom, we will also have to make the system size finite by imposing boundary conditions. One possible choice would be \emph{periodic} boundary conditions, so that the lattice is a discretization of the circle rather than the entire real line. These would require, for a linear arrangement of~$N$ total sites, that $\exp(ikNa) = 1$, i.e.\ that $k$ is an integer multiple of~$2\pi/Na$. (If one wanted to talk pretty, one would say that the translation symmetry is now $\Z/N\Z$, which is Pontryagin self-dual.) Of course, if the momenta of the mode we are interested in is large compared to the scale $2\pi/Na$ on which the discreteness occurs, the finite size doesn't make a substantial change in the behavior of the system. This should be familiar from everyday experience: the air in a room is obviously a system with finitely many degrees of freedom, in which the infrared (IR) regulator is the size of the room and the UV regulator is the intermolecular spacing. But the fact that one can speak words and be understood in such a room attests that the nondispersive wave equation is an exceptionally good model, at least for the wavelengths characteristic of the human voice.

Finite system size is often referred to as an \emph{infrared regulator} in physics, because it removes infinities that have to do with the momentum $k$ becoming infinitesimally small (or, what is the same, the wavelength of the mode becoming infinitely long) rather than large. As we have just seen, imposing both infrared and ultraviolet regulators typically yields a system with only finitely many degrees of freedom, whose behavior closely approximates that of the original system as long as the modes being studied are not close to either cutoff. Furthermore, as our example shows, the infrared cutoff typically takes the form of boundary conditions (periodic or otherwise), which can often be added or removed as desired without affecting the rest of the analysis or the solvability of the system. The limit $N \to \infty$ is referred to by physicists as the thermodynamic limit.

However, there is yet a third type of regulator that can be introduced, that pertains not to the number of degrees of freedom but to the nature of the degrees of freedom themselves. It is to this last approximation (and, with it, at last, to the subject of spin systems) that we turn in the next section.

\section{From field theories to spin systems}
\label{sec:FTtoSS}

So far, we've talked about field theories in which the degrees of freedom are functions on space, and we've shown how introducing a lattice discretization, together with a finite system size, can reduce a field theory's dynamics to that of only finitely many degrees of freedom, coupled in a homogeneous and local fashion. But each degree of freedom is still something like a harmonic oscillator; in particular, upon quantization, the Hilbert space of even one degree of freedom is still infinite-dimensional. 

We may be interested in replacing this problem by one in which the local Hilbert space is finite-dimensional. This has a number of advantages: First of all, it eliminates any remaining analytic issues. All operators are bounded and trace-class; they're just finite matrices. And one can in principle diagonalize such matrices---the Hamiltonian, for instance---on a computer, although (since the total dimension grows exponentially in the number of sites) this is infeasible in practice. As is usually the case, it's better to think structurally about what makes these matrices special than to attack the problem by brute force. And the relevant matrices are indeed a very special class, due once again to the pervasive notions of \emph{homogeneity} and~\emph{locality}.

What do homogeneity and locality mean for spin systems? Well, each of the sites is a subsystem, so it's clear that the total Hilbert space is a tensor product over all the sites:
\beq
\Hilb = \bigotimes_\alpha \Hilb_\alpha \qquad (\Hilb_\alpha \cong \C^d).
\eeq
There's therefore a notion of the \emph{support} of an operator: If it acts nontrivially on the tensor product of Hilbert spaces with $\alpha$ in some set~$S$, extended by the identity on all other sites as we discussed above, then its support is contained in~$S$. (The support is the intersection of all such~$S$.)
And if the dynamics are to be homogeneous and local, then we expect the Hamiltonian to be expressible as a sum over sites:
\beq
H = \sum_\alpha h_\alpha,
\eeq
where $h_\alpha$ does not depend explicitly on the parameter~$\alpha$, but rather is constructed in the same way at each site. Moreover, the support of~$h_\alpha$ should be ``local,'' in the sense that it includes $\alpha$ together with only a finite collection of nearby sites.

We promised that at least some systems satisfying these properties would admit clever techniques of solution, and this is indeed the case. Considering problems of this form opens the door to a wide class of exactly solvable models, often referred to as spin chains or quantum integrable systems. Integrability, of course, is an enormous and active field of research in its own right, and we will not have space to discuss it adequately here; the reader is referred to the literature (for instance, \cite{baxter,bethe,spinchains}). We will, however, give one important example of an integrable spin chain in the next section, which will hopefully give both some small idea of the flavor of integrable models and indicate the way in which such models are connected to field theories. In essence, the term means that such models admit reasonably simple closed-form expressions for the spectrum of the Hamiltonian as a function of parameters, and that the exact eigenstates can also be written down.

Why might a truncation to finite-dimensional local Hilbert spaces be reasonable or permissible to make in certain physical systems? Well, as is well-known, the quantum harmonic oscillator has a spectrum of energy eigenstates that are evenly spaced: 
\beq
H\ket{n} = E_n\ket{n} = \hbar \omega \qty( n+ \frac{1}{2} ) \ket{n}.
\eeq
If the Hamiltonians $h_\alpha$ of our lattice system consist, for example, of a harmonic oscillator Hamiltonian for the degree of freedom at~$\alpha$, together with terms that weakly couple it to its neighbors, then we know that the $\{ \ket{n_\alpha} \}$ (eigenstates of the uncoupled Hamiltonian) at least form a basis of the total Hilbert space. Of course, it won't be an energy eigenbasis anymore, but---since energy is a physically meaningful quantity that is stored locally by the degrees of freedom---we expect that states in which \emph{one} degree of freedom has an enormous amount of energy won't contribute meaningfully to states in which \emph{all} degrees of freedom collectively have much less energy. Thus, one imagines that eigenstates of the coupled system with energy less than some chosen bound can be constructed, at least to a very good approximation, as linear combinations of tensor product states where $n_\alpha$ is less than some cutoff value; and that is precisely the spin system approximation.

To emphasize the central point, this discussion indicates that the passage to a spin-system approximation of a field theory is best interpreted as a cutoff on the energy, or more precisely energy density, of the excitations we will allow ourselves to model. In terms of a field theory of light, we have already imposed a UV and an IR cutoff, that tell us not to think about wavelengths that are too large or too small; we expect this new approximation to work well (at least roughly speaking) only when the \emph{amplitude} of the light, or the number of photons present, is not too arbitrarily large.

Where does the name ``spin system'' come from? Well, the local Hilbert spaces are finite-dimensional, and the typical physical examples of systems with finite-dimensional Hilbert spaces are precisely spins. By Stone--von Neumann, any continuous degree of freedom has an infinite-dimensional Hilbert space, but the angular momentum algebra (which, again, is just~$\mathfrak{su}(2)$) admits a unique unitary representation of every finite dimension. So it is typical in physics to think of these Hilbert spaces as representing spins, whether or not there is actually $SO(3)$ invariance present.

What, then, is the classical analogue of a spin degree of freedom? 
For our purposes, we'll be satisfied with a simple (and of course very incomplete) answer, that the reader can probably guess from our discussion of quantization above. Since, for us, a ``spin'' is just a finite-dimensional Hilbert space, in which  a complete measurement produces only finitely many possible results, the corresponding space of classical states is just a finite set. Models made from such degrees of freedom (like the famous \emph{Ising model}, which we'll discuss more a bit later on) we will term \emph{classical spin systems}. Of course, any notion of classical dynamics as a Hamiltonian flow is out the window;
one typically has to interpret such models in the context of statistical mechanics at finite temperature. When the system is in thermal equilibrium with its environment, it effectively occupies each of its possible states $s$ at random with probability $p_s = \exp(- E_s/\tau)$. Here $\tau$ is a temperature parameter. We'll talk a bit more about systems of this kind below.
\begin{figure}
\begin{center}
\begin{tikzpicture}[xscale=1.3,yscale=1.5]
\tikzstyle{tbox}=[draw, text width = 2.15 cm]
\tikzstyle{over}=[preaction={draw, color=white, line width=6pt}]
\draw (0,0) node (A) [tbox]{classical field theory};
\draw (6,0) node (B) [tbox]{quantum field theory};
\draw (0,2) node (C) [tbox]{classical lattice system};
\draw (6,2) node (D) [tbox]{lattice field theory};
\draw (0,4) node (E) [tbox]{(semi)classical degree of freedom};
\draw (6,4) node (F) [tbox]{quantum \\ degree of freedom};
\draw (-3,3) node (G) [tbox]{finite set};
\draw (3,3) node (H) [tbox]{finite-dimensional Hilbert space};
\draw (-3,1) node (I) [tbox]{classical spin system};
\draw (3,1) node (J) [tbox]{quantum spin system};
\draw[->, double] (A.east) -- (B.west);
\draw[->, double] (C.east) -- (D.west);
\draw[->, double] (E.east) -- (F.west);
\draw[->, dotted] (C.south) -- (A.north);
\draw[->, dotted] (D.south) -- (B.north);
\draw[->] (E.south) -- (C.north);
\draw[->] (F.south) -- (D.north);
\draw[->] (G.south) -- (I.north);
\draw[->, over] (H.south) -- (J.north);
\draw[->, double, over] (G.east) -- (H.west);
\draw[->, double, over] (I.east) -- (J.west);
\draw[->, dashed] ([xshift=-6pt]E.south) -- ([yshift=+2pt]G.east);
\draw[->, dashed] ([xshift=-6pt]C.south) -- ([yshift=+2pt]I.east);
\draw[->, dashed] ([xshift=-6pt]D.south) -- (J.east);
\draw[->, dashed] ([xshift=-6pt]F.south) -- (H.east);
\end{tikzpicture}
\end{center}
\caption{A diagram of relations between different types of physical model}
\label{fig:bigCD}
\end{figure}
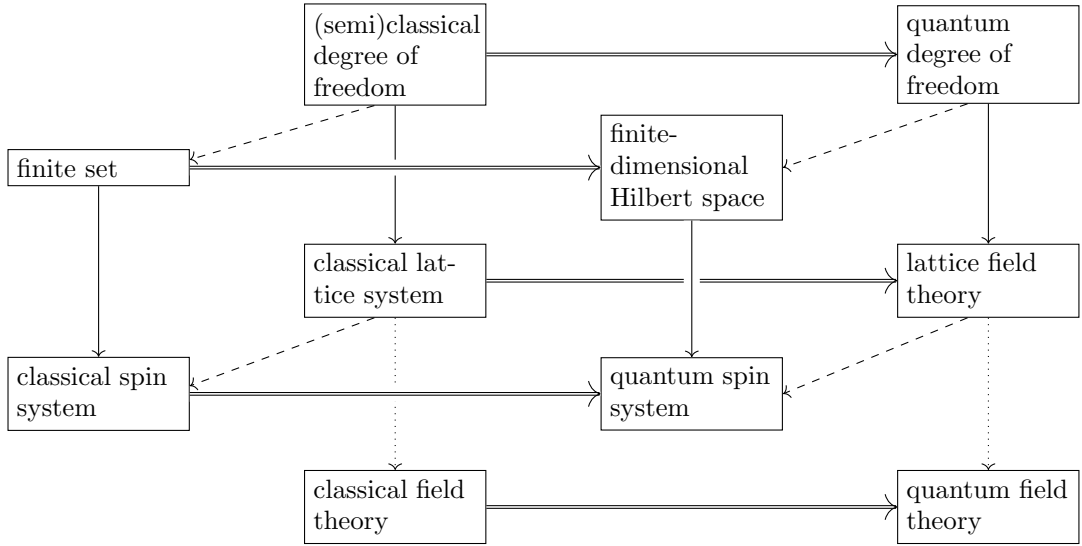

To summarize the discussion so far, and to keep the big picture in view, Figure~\ref{fig:bigCD} shows a cartoon of the relationships between the different types of physical systems we've been talking about. The top layer represents different choices of local degrees of freedom; the passage downward, as represented by arrows of the form  \arcoup, represents coupling many such degrees of freedom in a homogeneous, local manner.  Dashed arrows (\artrunc) represent a truncation of the kind we have discussed in this section, of the dimension of the local Hilbert space, and dotted arrows (\arcont) represent passage to a continuum limit. Finally, doubled arrows like \arquant\ represent quantization.

While we have chosen to represent the arrows with definite directions, we hope it is clear that the indicated relationships are somewhat more bilateral. 
In the discussion thusfar, we have taken a sort of ``top-down'' perspective, in which the field theory is viewed as the primordial object, and discretizations of it are approximations made to facilitate calculation or to remove problems with infinities. But we wish to emphasize that the reverse processes are always possible; 
that is, the discrete systems being considered \emph{behave like} continuum field theories, at least over a wide range of their possible excitations. 

The lesson one should draw is that what is truly fundamental is the picture of a field theory as \emph{many} degrees of freedom, coupled together in a way that is parameterized by a geometric space, so that the coupling is \emph{local} and \emph{homogeneous}. And quantum field theory is a universal language for talking about the collective behavior of such many-body systems. Indeed, in many physical applications (like the study of sound waves or magnetic materials), it is the discrete picture that is ``fundamental,'' and the continuum field theory that is a large-scale approximation! The lattice corresponds most closely to reality; field theory is nonetheless an exceptionally good model.

In some sense,  the real lesson is that what is ``fundamental'' is the wrong question to be asking. 
It is of little physical consequence whether ``many'' means $10^{23}$ or $\infty$, or even what the precise nature of each local degree of freedom is. But let's see how this plays out in another important example.


\section{An example: the Heisenberg spin chain and integrability}
\label{sec:XXX}

Following the criteria we've laid out, let's write down the simplest example of a quantum spin system we can, using the ingredients that have appeared in the preceding discussion. (Our discussion largely follows that in Bethe's original paper~\cite{bethe}, as well as~\cite{spinchains}.) We'll take the local Hilbert space to be two-dimensional; there is then a basis of local (one-site) Hermitian operators consisting of Pauli matrices acting at each site:
\beq
[ \sigma_i^\alpha, \sigma_j^\beta ] = i \epsilon_{ijk} \delta^{\alpha\beta} \sigma_k^\alpha .
\eeq
These generate (as an algebra) the entire collection of observables of the system.

The sites will be distributed along a one-dimensional lattice, just as in the discretized wave equation. 
We'll imagine that the degrees of freedom on each site are quantum spins with spin $S  =1/2$---i.e., two-dimensional Hilbert spaces. The mutual interaction between two such particles is a so-called \emph{exchange interaction}, which tends to align (or anti-align) the spins, and is represented in the Hamiltonian by a term proportional to $-\boldsymbol{\sigma}^\alpha\cdot \boldsymbol{\sigma}^\beta$---i.e., to the dot product of the two spin vectors. For more on the physics of exchange interactions and condensed matter systems more generally, see~\cite{sachdev}.

Furthermore, to ensure locality, we'll assume that the only significant exchange interactions occur between nearest neighbors in the lattice. 
We can now write a simple local translation-invariant Hamiltonian, satisfying all of the good properties we have outlined above: 
\beq
h_\alpha =  -\frac{J}{2} \qty( \boldsymbol{\sigma}^{\alpha-1} \cdot \boldsymbol{\sigma}^{\alpha}
+ \boldsymbol{\sigma}^\alpha \cdot \boldsymbol{\sigma}^{\alpha+1}
- \frac{1}{2} )
\eeq
For definiteness, we will assume that $J > 0$; such an exchange interaction is known as ferromagnetic.
We can, if we like, also add a term that represents an interaction with an external magnetic field to~$h_\alpha$, such as $B(\sigma^\alpha_z + 1/2)$. The total Hamiltonian then takes the form
\beq
H = \sum_\alpha h_\alpha = J \qty( \frac{N}{4} -  \sum_\alpha \boldsymbol{\sigma}^\alpha \cdot \boldsymbol{\sigma}^{\alpha+1} ) 
+ B \qty( \frac{N}{2}+ \sum_\alpha \sigma^\alpha_z ).
\label{eq:XXX}
\eeq
(The $N$-dependent constants are included merely to ensure that the ground-state energy is set to zero; they don't affect the physics, and one can safely ignore them). The Hamiltonian~\eqref{eq:XXX} encodes the dynamics of a simple integrable spin system, called the \emph{Heisenberg} or~\emph{XXX spin chain}. 

If we so desire, we can turn off the interactions by setting $J=0$. Then the system is just a composite of many noninteracting spins, and so is trivial to solve: the eigenstates are tensor product states of the form 
\beq
\ket{A} = \ket{a_1,\ldots,a_d} = \ket{\downarrow\downarrow\ldots\downarrow\uparrow\downarrow\ldots}.
\eeq
In this notation, $A$ is a subset of $\{\alpha\}$, the set of all lattice sites. It contains the positions of any spins that have been flipped up. The energy of such a state is
\beq
H \ket{A} = B\cdot{\#A}\cdot \ket{A}.
\eeq

To see that the interaction term changes the system meaningfully, we just need to check that the eigenstates of the free problem no longer solve the interacting Hamiltonian. And this is not difficult to do: 
 To see it simply, let's define $\sigma_\pm = \sigma_x \pm i \sigma_y$. These are raising and lowering operators; they satisfy the commutation relations
\beq
[\sigma_z, \sigma_\pm ] = \pm \sigma_\pm, \qquad [\sigma_+,\sigma_-] = 2 \sigma_z, 
\eeq
which should be familiar from the representation theory of~$\mathfrak{su}(2)$.
The Hamiltonian can then be rewritten in terms of these operators as
 \beq
 H = - J \sum_\alpha \qty( \sigma_z^\alpha \sigma_z^{\alpha+1} 
 	+ \frac{1}{2} \qty( \sigma_+^\alpha \sigma_-^{\alpha+1} + \sigma_-^\alpha \sigma_+^{\alpha+1} ))
	+ B \sum_\alpha \sigma_z^\alpha + E_0,
 \eeq
where $E_0$ denotes the (irrelevant) constant shift terms.
When one considers the action of this Hamiltonian on a tensor product state, it is clear that such a state (unless all spins are aligned) cannot possibly be an eigenstate: the terms containing~$\sigma_\pm$ represent ``hopping'' interactions, in which an up-spin moves one site right or left:
\beq
\stackMath
\sigma_-^\alpha \sigma_+^{\alpha+1}\ket{\ldots \downarrow\hsmash{\smash{\overset{\alpha}{\uparrow}}}\downarrow\ldots} = \ket{\ldots\downarrow\downarrow\uparrow\ldots}
\eeq
The action of $H$ on~$\ket{A}$ thus is a sum of terms in which all elements of~$A$ have been perturbed by~$\pm 1$. 
This expression for~$H$ also makes it obvious that~$H$ commutes with the operator $S_z = \sum_\alpha \sigma_z^\alpha$ that measures the total spin: none of the terms, including the hopping interactions, change the total spin of the state on which they act. As such, while the multi-index $A$ is no longer a good quantum number, its length $\#A$ (which is just the total spin) remains compatible with the interacting Hamiltonian. 

Indeed, if we were to set $B=0$, we could form the total spin operators
\beq
S_i = \sum_\alpha \sigma_i^\alpha,
\eeq
and \emph{all} of these would then commute with the Hamiltonian. Of course, they do not mutually commute; rather, they satisfy the $\mathfrak{su}(2)$ algebra. But the action of an algebra on~$\Hilb$, commuting with the Hamiltonian, is precisely a symmetry in quantum mechanics; this one comes from the obvious $SO(3)$ rotation symmetry of the space in which our spins live. Of course, this symmetry is broken by the application of an external magnetic field, as this picks out a distinguished direction. 

At this point, we've satisfied ourselves that~$H$ is an interesting, nontrivially interacting Hamiltonian, and we've also discovered that it commutes with (and so can be simultaneously diagonalized with)~$S_z$. We can therefore look for eigenvectors independently in each subspace with a definite value of~$\#A$. The subspace $\#A=0$ is already one-dimensional, so that state~$\ket{\emptyset}$ must be an energy eigenvector; in fact, it will be the vacuum state of the system. 

The next-easiest place to start is~$\#A = 1$. 
A general state in this subspace takes the form
\beq
\ket{\psi} = \sum_\alpha c_\alpha \ket{\alpha},
\eeq
and the eigenvalue equation we are trying to solve becomes
\beq
H\ket{\psi} = E\ket{\psi} = J \sum_{\alpha} c_\alpha \qty(
 \ket{\alpha} - \frac{1}{2} \ket{\alpha+1} - \frac{1}{2} \ket{\alpha-1} )
+ B \sum_\alpha c_\alpha \ket{\alpha}.
\eeq
A bit of rearrangement of terms puts this into the form of an equation for the $c_\alpha$'s:
\beq
(E-B) c_\alpha = J \qty( c_\alpha  - \frac{1}{2} c_{\alpha+1} - \frac{1}{2} c_{\alpha -1 }).
\label{eq:1mag}
\eeq
This should look surprisingly familiar; it is nothing other than the translation-invariant discrete equation we solved before, for the lattice wave equation! And, of course, this is no coincidence: it is the presence of discrete translation invariance that is responsible for the form of the solution in both cases. The solution, just as it was above, is to pick~$c_\alpha$ to be eigenfunctions $\exp(ik\alpha)$ of translation, reducing~\eqref{eq:1mag} to a dispersion relation:
\beq
E = B + J \qty( 1 -  \cos k ) = B + 2J \sin[2](k/2).
\label{eq:XXXdisp}
\eeq
Just as in the classical problem, we could choose to impose boundary conditions to make the problem genuinely finite. (We are then working in a subspace of dimension $N$, so of course there can be at most~$N$ possible eigenvalues of the energy; the allowed values of~$k$ must therefore be restricted whenever only finitely many lattice sites are present). A straightforward choice is periodic boundary conditions, which will impose the requirement that $\exp(ikN) = 1$, and thus that $k$ is an integral multiple of~$2\pi/N$, exactly as for the discrete wave equation. 

At this point, a physicist would say that we've shown that the Heisenberg spin chain admits quasiparticle-like excitations: just as the plane wave solutions in a field theory of electromagnetism correspond to photons of various wavelengths, so the plane-wave solutions we've just found correspond to collective excitations of definite wavelength, which (in this context) are sometimes called \emph{magnons}. Moreover, we've calculated their dispersion relation, which shows how the energy of a magnon is related to its wavelength. If we wanted to, we could write down ``magnon creation operators'' labeled by momenta:
\beq
a_k^\dagger = \sum_\alpha \exp(ik\alpha) \sigma_+^\alpha,
\label{magnon-eom}
\eeq
such that $a_k^\dagger \ket{\emptyset}$ is the ``one-magnon'' eigenstate with parameter~$k$ that we just constructed. 

Based on our experience above, we now have some idea of what to expect.  In \S\ref{sec:FT}, we started with a free, nondispersive model: the continuum wave equation. Discretizing led us to a free, but no longer nondispersive system (the discrete wave equation). Now, we've made a truncation in the local Hilbert-space dimension, and found a similar dispersion relation---even though we have no idea what the ``magnon equation of motion'' is. (Of course, we could guess it from~\eqref{magnon-eom}, observing that $\omega \propto k^2$ for long-wavelength modes.) 

Is the field theory of magnons still a free theory? If it were, we would have solved it completely; its Hilbert space would be just a Fock space, generated freely by the action of all the creation operators $a_k^\dagger$ on the vacuum. Moreover, the creation operators would all mutually commute with one another, and satisfy commutation relations like 
\beq
[H, a_k^\dagger] \stackrel{?}{=} E_k  a_k^\dagger,
\label{eq:conjCR}
\eeq
so that multi-magnon states like $a_k^\dagger a_{k'}^\dagger \ket{\emptyset}$ would be energy eigenstates. 


But this doesn't quite work. It is easy to check that, while the magnon creation operators mutually commute, they actually fail to commute with the Hamiltonian in the way specified by~\eqref{eq:conjCR}. 
One  computes that
\begin{align}
[H, a_k^\dag] &=
\sum_{\alpha, \beta} e^{ik\beta} \qty(
-J \qty( [\sigma_z^\alpha \sigma_z^{\alpha+1},\sigma_+^\beta]
+ \frac{1}{2} \qty( [\sigma_+^\alpha \sigma_-^{\alpha+1},\sigma_+^\beta] + [\sigma_-^\alpha \sigma_+^{\alpha+1},\sigma_+^\beta] )
) + B [ \sigma_z^\alpha, \sigma_+^\beta ]
) \nonumber \\
&= \sum_{\alpha, \beta} e^{ik\beta} \qty(
-J \qty( 
\sigma_z^\alpha \sigma_+^{\alpha+1} \delta^{\alpha+1,\beta}
+
\sigma_+^\alpha \sigma_z^{\alpha+1} \delta^{\alpha\beta}
-
\sigma_+^\alpha \sigma_z^{\alpha+1} \delta^{\alpha+1,\beta}
-
\sigma_z^\alpha \sigma_+^{\alpha+1} \delta^{\alpha\beta}
)
+B \delta^{\alpha\beta} \sigma_+^\beta )
\nonumber \\
&= B a_k^\dag - J \sum_\alpha e^{ik\alpha} \qty(
\sigma_z^\alpha \sigma_+^{\alpha + 1}
-
\sigma_+^\alpha \sigma_z^{\alpha + 1}
)
\qty(
e^{ik} - 1
).
\label{eq:Hcomm}
\end{align}
This does not vanish, and so one can't expect that ``multiparticle states'' created by several successive magnon creation operators are energy eigenstates. Indeed, the only reason that single-particle states are energy eigenstates is that the operator appearing next to~$J$ in~\eqref{eq:Hcomm} 
acts simply on the vacuum. 
The $\sigma_z$ operators all act by $-1/2$, so that one obtains
\beq
[H, a_k^\dag] \vac = B a_k^\dag \vac + \frac{J}{2} (e^{ik}-1)(e^{-ik}-1) a_k^\dag \vac = 
\qty( B + 2J \sin[2](k/2) ) a_k^\dag \vac,
\eeq
recovering the single-particle dispersion relation, as of course one must.





Remarkably, though, one can recover the correct two-magnon eigenstates by a small, but clever, modification of the state $a^\dag_k a^\dag_{k'} \vac$. The form of the resulting solution goes by the name of the \emph{Bethe ansatz}, and plays a key role in the story of integrable systems. One arrives at it by remembering that, in our computation~\eqref{eq:Hcomm}, the problematic term (which we want to reduce to be proportional to~$a_k^\dag$) in fact \emph{does} work out correctly, as long as the two spins are separated from one another. The problem only appears when the spins are neighbors. 

Let's write out the state as follows:
\beq
a^\dag_k a^\dag_{k'} \vac = \sum_{\alpha\neq \beta} e^{i(k\alpha+ k'\beta)} \ket{\alpha\beta} = \sum_{\alpha<\beta} \qty(e^{i(k\alpha + k'\beta)} + e^{i(k'\alpha + k\beta)}) \ket{\alpha\beta}.
\label{eq:2free}
\eeq
Bethe's insight was to consider the slightly more general state
\beq
\sum_{\alpha<\beta} \qty(e^{i(k\alpha + k'\beta)} + Ae^{i(k'\alpha + k\beta)}) \ket{\alpha\beta},
\label{eq:2BA}
\eeq
where~$A$ is a constant to be determined. One imagines that it represents an interaction of some sort, that takes place when one magnon passes through the other. By using this ansatz, it is a good exercise to check that~\eqref{eq:2BA} gives an eigenstate of the Hamiltonian precisely when 
\beq
A = \frac{\cot(k/2) - \cot(k'/2) - 2 i}{\cot(k/2) - \cot(k'/2) + 2 i} = \exp i \theta(k,k').
\eeq
$A$ is thus a pure phase, which is a function only of the two magnon momenta. It encodes magnon-magnon scattering. One is led to the conclusion that the particles are  \emph{almost} free: they can pass through one another without exchanging any momentum, but, in doing so, they acquire a phase shift. In the time domain, the meaning of this is that they experience a momentum-dependent time delay.

The energy of the resulting state is just the sum of the energies of the two magnons independently; that is, $E(k) + E(k')$, with~$E$ defined using the dispersion relation~\eqref{eq:XXXdisp}. One might therefore be tempted to say that the interaction has no effect on the energy spectrum whatsoever. But this isn't quite true! When periodic boundary conditions are imposed, the quantization conditions on allowed values of~$k$ and~$k'$ are altered by the scattering. They become
\beq
kN = \theta(k',k),  \quad k'N = \theta(k,k')  \pmod{2\pi}.
\eeq
And this means that the allowed \emph{discrete} spectrum of two-magnon states isn't quite just the set of sums of allowed energies of one-magnon states.

The truly amazing thing is that Bethe's ansatz works to find exact eigenstates when \emph{any} number of quasiparticles are present. One simply writes down the ``free-theory'' ansatz analogous to~\eqref{eq:2free}, writing it as a sum over basis vectors with each coefficient carrying terms from all possible reorderings of the momenta.  Then, one adds a phase for each pairwise interaction between particles. The result is an exact multi-magnon state, and the quantization conditions are also simple to intuitively understand. They just state that the wavefunction for each magnon must be single-valued, after accounting for the phase acquired by transporting it past all the others:
\beq
k_i N = \sum_{j\neq i} \theta(k_j,k_i) \pmod{2\pi}. 
\label{eq:quant}
\eeq
One can then sum the single-magnon energies for any collection of $k_i$ permitted by the quantizability conditions, and obtain the corresponding energy eigenvalue of the system. Remarkably, the Bethe ansatz produces a \emph{complete} solution of the model, with a pleasing physical interpretation, in terms of simple algebraic constraints~\eqref{eq:quant} on parameters!

It would be easy to fill this volume (and several more besides) with material about integrable systems. 
So we'll have to make an abrupt exit from the subject at some point, and this is as good a time as any. 
Given just a bit more space, one would like to discuss how integrable models are reformulated in terms of an \emph{algebraic} version of the Bethe ansatz, which constructs a Lax representation of the Hamiltonian and its conserved quantities, and encodes the scattering data $\theta$ in terms of an $R$-matrix satisfying the famous Yang--Baxter equation.  
Such $R$-matrices are closely connected to braid group representations, and therefore to the theory of anyons in two-dimensional systems. The connection continues on to knot invariants, and hence to line operators in three-dimensional TQFT (another theme of ours). But we will have to leave this bridge unbuilt, and refer instead to the extensive literature on the subject~\cite{froehlich, turaev}.

By way of closing, let us just mention that we earlier defined complete integrability as the existence of as many mutually compatible conserved quantities as possible, and suggested that it is often connected to the action of a symmetry algebra on the system (for example, rotational invariance in the hydrogen atom). These ideas hold true for the Heisenberg chain, which admits not just the $\mathfrak{su}(2)$ algebra we defined before, but a much larger symmetry algebra. 

We'll see later, in our discussion of the toric code model, that the truly crucial feature is the presence of an algebra of non-local symmetries (which in that case are line operators). Non-local symmetries are also a crucial theme in the study of integrable systems: the additional symmetries of the Heisenberg chain act non-locally, and together constitute an algebra called the Yangian.
For example, the quadratic generators of this algebra take the following form:
\beq
S_z^{(2)} = \sum_{\alpha < \beta} (\sigma_+^\alpha \sigma_-^\beta - \sigma_-^\alpha \sigma_+^\beta), \qquad
S_{\pm}^{(2)} = \sum_{\alpha < \beta} (\sigma_z^\alpha \sigma_\pm^\beta - \sigma_\pm^\alpha \sigma_z^\beta).
\eeq
It's an instructive exercise to show that these operators commute with the Hamiltonian, as long as periodic boundary conditions are imposed. Further commutation relations of these operators generate new, higher-order nonlocal symmetries; the Yangian is the algebra of all of these taken together. It is a deformation of the algebra of polynomials with coefficients in (the universal enveloping algebra of)~$\mathfrak{su}(2)$. Further discussion of Yangian symmetry can be found, for example, in~\cite{costello,yangian}.


\section{Questions to ask about systems, or, What is a phase of matter?}

Having spent so long discussing different types of physical models, and the way in which they all arise from field theories by successive processes of discretization or regularization---or, to take the complementary perspective, illustrating the way in which field theory is a universal language for talking about a wide class of discrete and continuous models in a regime of their parameter space, provided that they consist of many similar degrees of freedom coupled in a spatially local and homogeneous way---we will now turn and say a few words about the kind of questions physicists might ask about the behavior of such systems. The kind of thing one would like to do
is to make a kind of field guide to such systems, or otherwise understand the ``zoology'' of possible behaviors they can exhibit. 
To have some clear ways to start classifying them, one wants a list of well-defined and salient questions (invariants or characteristic features) that can tell them apart. 


As we've already emphasized, field theory should primarily be thought of in a physical context as an effective or emergent description, rather than something complete and self-contained. One often doesn't know a precise description of the microscopic physics underlying the particular effective field theory one is studying, and (especially in applications to condensed matter) it is always possible that various impurities and additional interactions are present as perturbations to the system. 
So what is most often interesting about a field theory is its \emph{infrared} behavior, i.e., how it behaves for modes of sufficiently long wavelength or low energy. Furthermore, one would typically only think of a characteristic feature or behavior of the theory as meaningful if it is robust against small perturbations or variations of the parameters of the system. 

A first-order question of this kind would be to simply ask about the spectrum of the Hamiltonian. In general, this isn't a wildly interesting problem in field theories; since particle-like solutions typically exist with variable momenta, the spectrum of the Hamiltonian is usually continuous, at least above a certain value of the energy. But the low-lying spectrum can differ: theories exist in which the continuous spectrum extends all the way to the lowest-energy state. These are termed \emph{gapless} theories. On the other hand, theories may also be \emph{gapped:} that is, the vacuum may be in a discrete part of the spectrum, separated by some amount~$\Delta$ from any other states, whether continuous or discrete. (A typical example is a field theory in which all particles are massive; the gap is then the mass of the lightest particle.) It is clear that the set of gapped theories is ``open,'' in that an infinitesimal perturbation to a gapped system will result in another gapped system. Put differently, the property of being gapped is robust, in the sense we discussed above.

Of course, in a spin system, the Hilbert space is finite-dimensional, and so no portion of the spectrum can actually be continuous. So we need to interpret these ideas in the context of a thermodynamic limit. That is, we say a system is gapped if the gap persists at a finite value in the limit where the number of lattice sites grows to infinity. 

We've already decided that one should consider small variations to the parameters of the Hamiltonian as insignificant from the point of view of the classification we'd like to achieve. This means that two Hamiltonians $H_0$ and~$H_1$ should be considered equivalent if they're connected by a sufficiently short path~$H_t$ in the space of possible Hamiltonians. But what should ``sufficiently short'' mean? In the context of gapped theories, there's an immediate answer: any such path between gapped Hamiltonians can be subdivided into arbitrarily ``short'' equivalences between nearby gapped Hamiltonians, \emph{unless} the gap closes at some point along the path. So the natural thing to consider is the set of path components of the collection of gapped Hamiltonians. We call such an equivalence class of gapped systems a \emph{phase}.

How can one characterize phases? Well, the next simple thing one might ask is about a gapped theory is whether or not its vacuum state is unique. 
 As we will discuss, in certain situations the vacuum degeneracy is an invariant of the phase. So one is led to ask about what kinds of degenerate vacua can appear in field theories or spin systems. This, of course, naturally suggests that one should study TQFT---but what kinds of field theories should one expect to have vacuum degeneracies described by interesting TQFTs?

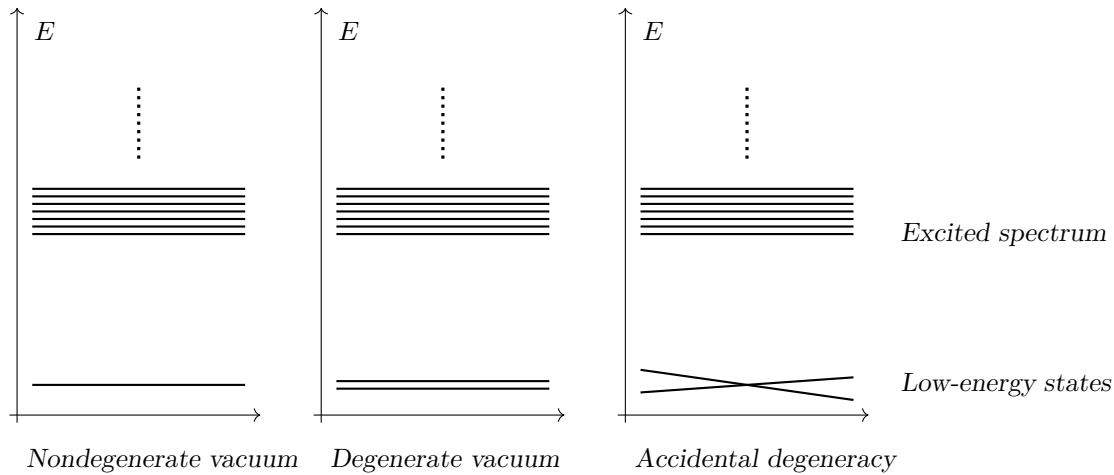
\begin{figure}
 \begin{center}
 \begin{tikzpicture}
 \tikzstyle{tbox}=[anchor=west] 
 \foreach \x in {0,4,8} {
	 \draw[->] (\x,-0.5) -- (\x,5);
	\draw[->] (\x - 0.1,-0.4) -- (\x + 3.2,-0.4);
	 \draw (\x +0.1,4.7) node[anchor=west]{$E$};
 	\foreach \y in {0,0.1,0.2,...,0.6} {
		\draw[thick] (\x+0.2,2+\y) -- (\x+3,2+\y);
		};
	\draw[dotted, very thick] (\x+1.6,3) -- (\x+1.6,4); 
	};
 \draw[thick] (0.2,0) -- (3,0);
 \draw[thick] (4+0.2,-0.05) -- (4+3,-0.05);	
\draw[thick] (4+0.2,0.05) -- (4+3,0.05);
\draw[thick] (8.2,-0.1) -- (11,0.1);
\draw[thick] (8.2,0.2) -- (11,-0.2);
\draw (11.5,0) node[tbox]{\sl Low-energy states};
\draw (11.5,2) node[tbox]{\sl Excited spectrum};
\draw (0,-1) node[tbox]{\sl Nondegenerate vacuum};
\draw (4,-1) node[tbox]{\sl Degenerate vacuum};
\draw (8,-1) node[tbox]{\sl Accidental degeneracy};
 \end{tikzpicture}
 \end{center}
 \caption[Different possible situations in gapped systems.]{Different possible situations in gapped systems. The abscissa indicates a parameter, or possible path of nearby Hamiltonians.}
 \label{fig:gaps}
 \end{figure}

As we mentioned earlier, degeneracies in quantum-mechanical systems are generically not present. It is often the case that the Hamiltonian alone will constitute a complete set of observables; in a family of generic Hamiltonians, one expects degeneracies in the energy spectrum to occur only on a locus of positive codimension.  Such degeneracies are sometimes termed \emph{accidental}; for an illustration, see Fig.~\ref{fig:gaps}, at right.
So then how can robust ground state degeneracy occur? One mechanism involves the presence of a symmetry and the phenomenon of \emph{spontaneous symmetry breaking}.

As we have discussed above, a symmetry is present in a quantum-mechanical system when the Hamiltonian commutes with the action of some algebra. This may be an algebra of observables, corresponding to the generators of a continuous symmetry, or simply the group algebra of a discrete group of symmetries. Such a situation implies that each eigenspace of the Hamiltonian carries a representation of the algebra. It might happen that the ground state in the thermodynamic limit carries a nontrivial representation; 
this is termed a spontaneous breaking of the symmetry. 
The terminology arises because the vacuum fails to be invariant under the symmetry, even though the Hamiltonian that determines it \emph{is} invariant. (In a sense, the solutions to the equations of motion are less symmetric than the equations of motion themselves.)
When the representation on the vacuum is of dimension larger than one, spontaneous symmetry breaking results in a vacuum degeneracy.

Note that while in a finite system each eigenstate will generally carry an irreducible representation (absent accidental degeneracy), in the thermodynamic limit several irreducible representations might become degenerate in energy. In particular, the ground state might carry a reducible representation.




The prototypical phenomenon to keep in mind here is something like ferromagnetism in iron. Iron consists of a number of constituent molecules with a spin, so that each produces a magnetic field and interacts with its neighbors in a manner similar to the Heisenberg model discussed above. At high temperatures, the equilibrium state of the iron will be disordered; the interactions will be insufficient to align faraway spins, and the net magnetization will be zero. As the system is cooled, though, the role of interactions becomes more important, and below a particular critical temperature the equilibrium state of the iron will  be a permanent magnet. The spatial rotation symmetry is then spontaneously broken: the magnetization points in a particular direction in space, but the rotation group acts to produce other inequivalent equilibrium states.

One can therefore detect what phase the system is in by measuring the magnetic field: when its magnitude is zero, the system is in the disordered phase (the symmetry is unbroken). Conversely, when it is nonzero, the symmetry is broken and the system is in the spontaneously ordered phase. Observables that detect the phase of the system such as this are termed \emph{order parameters;} it is clear from our discussion that the order parameter necessarily exhibits nonanalytic dependence on the temperature at the phase transition. Landau proposed a theory of phase transitions with precisely these ingredients: a symmetry that becomes spontaneously broken, and a locally measurable order parameter that detects the corresponding phase transition. Although simple, the Landau theory was extremely successful, predicting even the precise kind of nonanalytic behavior one should expect of the order parameter near the phase transition.

A simple but famous semiclassical model of ferromagnetic behavior\footnote{For a discussion of an analogous quantum-mechanical model,  the transverse field quantum Ising model, see, for example, \cite[chapter 5]{sachdev}.} is the \emph{Ising model}, which we mentioned briefly above. In this model, one imagines that the classical set of states is defined by a collection of spins taking the values $\pm 1$; these interact with their nearest neighbors in a way that favors alignment. The system is then taken to occupy each of its states at random, with probability given by $\exp(-E/\tau)$. One thus sees that there is a competition between order and disorder. States in which all spins are aligned are the energetically most favorable, and so the most probable. On the other hand, there are enormously more states in which some number of spins are misaligned; for example, the degeneracy of a state with half spins up and half down is ${N\choose N/2}$, which is an incredibly large number. If the temperature is taken to  infinity, all states are equally likely, so that disorder clearly triumphs. On the other hand, at zero temperature, any excitation has zero probability, so that only the ordered state is possible. That there is a sharp phase transition between these two phases in the two-dimensional Ising model is a celebrated result of Peierls (the system was later solved exactly by Onsager). 

The equilibrium state in a thermodynamic system like the Ising model can be determined by minimizing the \emph{free energy}, which is defined to be the combination $F = E - \tau S$. Here $E$ is the energy and $\tau$ the temperature, as before, while $S$ is the entropy, which measures disorder, or more precisely the number of different degenerate states available to the system. In Landau's theory, one writes the free energy as a function of the order parameter (here, the magnetization $\phi$), and postulates that it is both analytic and invariant under the relevant symmetry. The typical model for a phase transition is then of the form
\beq
F(\phi) =  (\tau - \tau_c) \phi^2 + \lambda \phi^4 + \ldots,
\label{eq:SSB}
\eeq
where the dots represent possible higher-order terms in a Taylor series. (Here we are treating the simplest case, where $\phi$ is valued in~$\R$, and the relevant symmetry is $O(1) \cong \Z/2\Z$.) One sees immediately that, even though $F$ is perfectly analytic  in $\phi$ and $\tau$, the equilibrium value $\phi_0$ of the  order parameter (i.e. the minimum of~$F$) is not: rather,
\beq
\phi_0 = 
\begin{cases}
\sqrt{
{(\tau_c - \tau)}/{\lambda}}, & \tau < \tau_c; \\
0, & \tau > \tau_c.
\end{cases}
\eeq
Landau's simple analysis of the phase transition can be refined to include fluctuations of the order parameter; the complete framework is known as the Landau--Ginzburg--Wilson paradigm of classical phase transitions.
It is worth noting that the example~\eqref{eq:SSB} famously recurs in elementary particle physics, in the context of spontaneously broken gauge symmetry. It is the self-interaction potential of the Higgs boson.

Successful as the Landau--Ginzburg--Wilson theory was, it was eventually realized that it did not provide a perfect classification of phases. Indeed, physicists found examples of systems in which \emph{no} local symmetry breaking or order parameter accounted for the degeneracy. Let us say that two degenerate vacua, $\ket{a}$ and~$\ket{b}$, are \emph{locally indistinguishable} if, for any local operator~$\hat{A}$,
\beq
\bra{a}\hat{A}\ket{b} \propto \delta_{ab}.
\eeq
That is, $\hat{A}$ acts in the space of vacua by an operator proportional to the identity, and so measuring it does not serve to distinguish one degenerate vacuum from another.

The study of phases with locally indistingushable robust vacuum degeneracy has given rise to the subject of \emph{topological order}, which is more or less the topic of the conference. Physicists have had to go beyond Landau's theory of phases; in doing so, they have taken seriously the idea that the structure that controls the degenerate vacuum sector or deep infrared limit of a topologically ordered system is a nontrivial topological field theory. (For early papers on the subject that are still well worth reading, see~\cite{wen1,wen2}.) In the coming section, we'll say a bit more about phases, the structures that they have, and what is known about them. Then we'll conclude with an example of a spin system with topological order, that will shed some light on how topologically ordered systems fit into the idea that robust degeneracies only arise from the action of a symmetry. In this instance, that ``symmetry'' will be related to the presence of \emph{anyons}  (particles with unusual exchange statistics).
Indeed, topological quantum field theories in spacetime dimension three can all be thought of as  arising from  anyons~\cite{rowell-wang,turaev2}. And one can even often construct Hamiltonians whose vacuum sector represents such a TQFT: a construction of Levin and Wen~\cite{levin-wen}, generalizing the toric code model, produces such a model from the data of a modular tensor category. (The Levin-Wen construction only gives a subset of topological orders, those with ``gappable boundaries,'' but does produce all parity- and time-reversal-invariant topological phases; see~\cite{levin,levin-wen} for details, as well as~\cite{kong,kitaev-kong} for connections to extended TQFT.) 

We'll see why ground state degeneracies are connected to the presence of anyons a bit later on; to foreshadow a bit, the essential idea is that \emph{nonlocal} operators can be constructed from the worldlines of anyons, which act as an algebra and commute with the Hamiltonian. This algebra is then ``spontaneously broken,'' resulting in a nontrivial robust vacuum degeneracy. In some sense, therefore, Landau's paradigm is still intact; one just needs to relax the assumption that the symmetry algebra consists of local operators. Versions of this idea go back to 't~Hooft, who introduced line operators as order parameters in gauge theory~\cite{thooft}; see~\cite{generalized} for a recent discussion of generalized non-local symmetry.

It's also worth mentioning here that there are slightly more general notions of phase that one might consider, defined by cutting down the class of Hamiltonians one is interested in. For instance, one might insist that all Hamiltonians be time-reversal invariant, or respect some other discrete symmetry. Then the relevant question becomes to understand gapped \emph{symmetric} Hamiltonians, regarded as equivalent if they can be connected by a path of gapped symmetric Hamiltonians. Notions like this are termed \emph{symmetry-protected topological order} in the literature, but we won't have space to offer further discussion here; the reader is referred to the other lectures in this volume. We'll also have to give a slight refinement of our definition of a phase, that allows one to change the dimension of the local Hilbert space; we turn to this in the next section.

\section{Stacking}

A perspective that we've tried to highlight throughout this lecture is the crucial importance of the notion of composition of subsystems in physical theories. Other fundamental ideas, such as locality, only make sense if we know what it means for a system to be composed of subsystems and what it means for two subsystems to be noninteracting. 

The reader can likely guess that there's a sense in which two separate field theories (or spin systems), living over the same space (or lattice), can be concatenated as well. This operation is sometimes called ``stacking,'' and it's defined in the obvious way: The new local Hilbert space is the tensor product of the local Hilbert spaces of each system; the new local Hamiltonians $h_\alpha$ are the sums of the previous local Hamiltonians, extended by tensoring with the identity. 

This gives a notion of composition on spin systems, making the collection of spin systems into an abelian monoid. How does this play with the notion of equivalence we defined before? Well, it's clear that an equivalence between systems $A$ and $A'$ induces an equivalence between $A\otimes B$ and $A' \otimes B$, and likewise for the other subsystem. So stacking descends to a composition law on phases. One thing to note, though, is that this composition is actually valued in a more fine space that maps down to the space of phases. Any equivalence that varies the parameters of only one subsystem is a path of \emph{interaction-free} Hamiltonians of the composite. There are thus two questions one might ask about a spin system that is the composite of two constituent spin systems: one might ask about the path-components of the set of interaction-free Hamiltonians, or about the path-components of all Hamiltonians (ignoring the decomposition). The first maps to the second, but there is no  reason the map must be either injective or surjective: there may be phases that contain no representative noninteracting Hamiltonian, and two noninteracting Hamiltonians may be connected by a path only when interactions are allowed to be turned on. 

As the reader can probably guess, analogous considerations imply that phases are classified differently depending on whether or not one considers interacting systems, or just free systems---i.e., whether one considers the path components of the space of free Hamiltonians, or all Hamiltonians. Again, this map is in general neither injective nor surjective. For a recent discussion, see~\cite{KTTW}.

There's also a sensible physical notion of a trivial or identity phase: we'll say that a Hamiltonian represents the trivial phase if it has a unique ground state, which is a tensor product over all local degrees of freedom. That is, it's of the form
\beq
\vac = \bigotimes_\alpha \ket{\psi}.
\eeq
Put differently, the local degrees of freedom in a trivial phase don't have any meaningful interaction (or correlation, or entanglement) with one another at all. 

It would be nice if the ``identity'' phase were the identity for our composition operation, but that's not true as we've defined things so far. The essential problem is that, as we emphasized before, the local Hilbert space dimension is not an essential detail or ingredient, and different systems with different values of the local dimension may represent the same physics. But our definition of a phase only allows us to regard systems as equivalent when the local dimensions agree. 

To remedy this, let's rectify our definition by making a notion of ``stable equivalence:'' We'll say that two spin systems are equivalent if they can be connected by a path of Hamiltonians, after either of them is tensored with any number of Hamiltonians representing the trivial phase. The trivial phase is then the identity for stacking by definition. 

Now we have a unital abelian monoid of phases, and so we can ask what the group of \emph{invertible} phases is. These are Hamiltonians~$H$ representing phases that are nontrivial, but for which a partner Hamiltonian $\bar{H}$ can be found such that $H+\bar{H}$ admits a deformation to the trivial phase. Computing these groups of invertible phases, possibly with additional subtleties such as protecting symmetry groups, is the essential subject of Dan's lectures, and we leave it to him to discuss; a table showing a couple of examples of groups of invertible phases is given in Fig.~\ref{fig:invph}.

\begin{figure}
\begin{tabular}{c|c|c|c}
\emph{spacetime dim.} & 2 & 3 & 4 \\ \hline
bosons & 0 & $\Z$ & 0 \\
& & $\nu = 8$ integer QHE & {\qquad\qquad} \\ \hline
fermions & $\Z/2\Z$ & $\Z$ & 0 \\
& Majorana wire & $p+ip$ superconductor & {\qquad\qquad} \\
& & ($\nu=1$ QHE represents $2\in\Z$) & \\
\end{tabular}
\caption{Groups of invertible phases  under stacking in various dimensions.}
\label{fig:invph}
\end{figure}

%






\section{Topological order and the toric code}

As a last invitation, we'll discuss a very famous spin system, due to Alexei Kitaev~\cite{kitaev}, that exhibits a protected ground-state degeneracy \emph{without} a local order parameter. This model goes by the name of the \emph{toric code;} while it will be discussed again elsewhere (e.g.~in Dan Freed's lectures in this volume), we feel it is useful to give a physicist's perspective on the model and its features of interest, and to indicate how it fits into the bird's-eye view we have been outlining. 
Our presentation is slightly nonstandard, but we feel that it is the clearest and simplest way to analyze the system. 

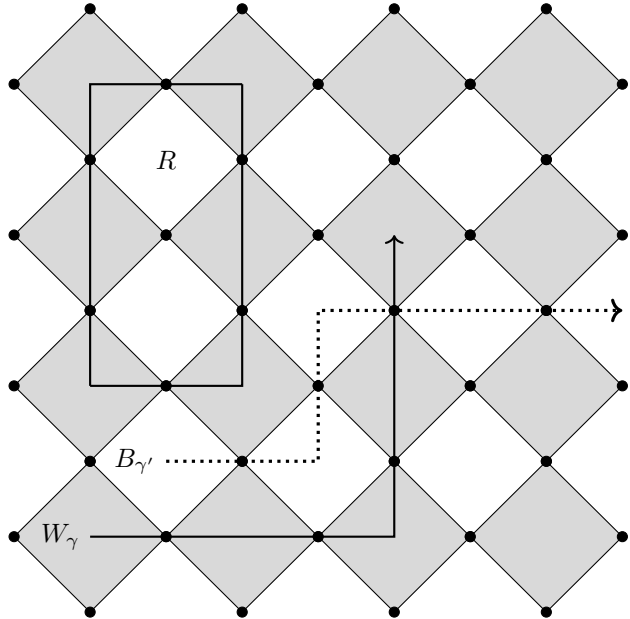
\begin{figure}
\begin{center}
\begin{tikzpicture}
\tikzstyle{site}=[draw,scale=0.4,fill=black,circle]
\foreach \y in {0,2,4,6} {
	\foreach \x in {0,2,4,6} {
		\draw[fill=gray!30!white] 
			(\x,\y) node[site]{} -- 
			(\x+1,\y+1) node[site]{} -- 
			(\x+2,\y+0) node[site]{} -- 
			(\x+1,\y-1) node[site]{} -- 
			cycle;
		};	
	};
\draw[dotted, very thick, ->] (2,1) node[anchor=east]{$B_{\gamma'}$} -- (4,1) -- (4,3) -- (8,3);
\draw[
thick, ->] (1,0) node[anchor=east]{$W_{\gamma}$} -- (5,0) -- (5,4);
\draw[thick] (1,2) -- (1,6) 
	-- (3,6);
\draw[thick] (3,6) -- (3,2) -- (1,2);
\draw (2,5) node{$R$};
\end{tikzpicture}
\end{center}
\caption{A picture of the toric code and its line operators}
\label{fig:toric}
\end{figure}

The underlying lattice of the toric code is a two-dimensional square lattice, which we'll think of as being bicolored with alternating black and white squares like a chessboard, and oriented diagonally, as shown in Fig.~\ref{fig:toric}. Just as in the Heisenberg model, each vertex of the lattice carries a local Hilbert space, which is taken to be two dimensional. 
The algebra of observables is therefore again generated (as an algebra) by the one-site Pauli matrices $\sigma_i^\alpha$. 

Consider the following operators, defined to act on the four sites surrounding either a black or a white square: 
\beq
h_w = \prod_{\alpha \in w} 2\sigma_x^\alpha, \qquad 
h_b = \prod_{\alpha \in b} 2\sigma_z^\alpha.
\label{eq:sqHams}
\eeq
(In what follows, the label $w$ will always refer to a chosen white square, and~$b$ to a black one.) The factors of two are included because of the normalization; for our purposes here, it is important that $(2\sigma_i)^2 = 1$, for reasons that will become clear momentarily. 

It is obvious that all of the black-square and all of the white-square operators mutually commute. What is slightly less obvious is that in fact \emph{all} of the operators~\eqref{eq:sqHams} are mutually commuting. The case to check is that of a white square and a black square, adjacent along an edge. The important calculation (which makes a straightforward but healthful exercise) is
\beq
[\sigma_x^\alpha \sigma_x^\beta, \sigma_z^\alpha \sigma_z^\beta] = 0.
\eeq
From this, it follows that all of the $h_w$ and~$h_b$ can be simultaneously diagonalized! Since they square to the identity, each of these operators has possible eigenvalues $\pm 1$; however, because the product over all~$w$ of~$h_w$ is the identity operator (and similarly for the~$h_b$), there are two constraints on the set of the mutual eigenvalues, which insist that the number of white squares $w$ with~$h_w=-1$ is even, and likewise for black squares. (In the end, we will choose periodic boundary conditions.) 

The Hamiltonian of the toric code is just a linear combination of all these mutually commuting local Hamiltonians:
\beq
H = - \sum_w h_w - \sum_b h_b.
\eeq
We have already, in some sense, computed the spectrum of this Hamiltonian: we expect that all the local operators $h$ can be simultaneously diagonalized, and any set of eigenvalues chosen that obey the constraints we mentioned above can be realized. But let's try to get a clearer picture of the behavior of the model, and (in particular) figure out whether or not the ground state is degenerate. 

To do this, one should consider defining the following \emph{line operators:}
\beq
W_{\gamma} = \prod_{\alpha \in \gamma'} 2\sigma_x^\alpha, \qquad
B_{\gamma'} = \prod_{\alpha \in \gamma} 2\sigma_z^\alpha.
\eeq
Here, the paths $\gamma$ and~$\gamma'$ with respect to which the operators are defined, are paths of squares like those taken by chess bishops: they pass over squares of the same color, always moving diagonally. A site $\alpha$ belongs to a path if the colored squares on both sides of it do, i.e., if a chess bishop moving along the path of squares would have passed over that vertex. The picture in Fig.~\ref{fig:toric} should make this clear.
A word of caution: the operators $W$ are defined on paths corresponding to chess bishops on the \emph{black} squares, and $B$ on those on the \emph{white} squares! The reversal occurs because the path of a chess bishop on white squares defines a partition of the black squares, and vice versa; a clash in the notation in some place cannot be avoided.\footnote{Of course, the color of the bishop doesn't correspond to the color of the squares on which it moves in chess either.}

When~$\gamma$ is taken to be the shortest possible closed path, so that it has length four, turns in a square and encloses exactly one white square, it is easy to see that $W_\gamma = h_w$, where~$w$ is the enclosed square. $W_\gamma$ therefore commutes with all local Hamiltonians. Indeed, this generalizes to \emph{any} closed path: since $\sigma_i^2 = 1$, the product of two adjacent operators $h_w h_w'$ is equal to the line operator on the closed path of length six encircling both adjacent squares. (See the region marked~$R$ in Fig.~\ref{fig:toric}.) And it is trivial to see that this generalizes to
\beq
W_{\bdy R} = \prod_{w \in R'} h_w, \qquad
B_{\bdy R'} = \prod_{b \in R} h_b,
\eeq
for any regions $R$, $R'$ and corresponding nullhomologous boundary curves.

What if the curves are open? Well, the easiest way to understand this case is to look at the commutation relations between different line operators. For this, it's useful to recall that the Pauli matrices also obey anticommutation relations of the form 
\beq
\{ 2\sigma_i^\alpha, 2 \sigma_j^\beta \} = 2 \delta^{\alpha\beta}\delta_{ij}.
\eeq
Now, it's easy to see that all the $W$ operators commute with one another, independent of their support (all the local operators they're built out of mutually commute), and likewise for the $B$ operators. But what about a $W_\gamma$ and a $B_{\gamma'}$? Well, suppose that the paths $\gamma$ and~$\gamma'$ cross exactly once, at a site~$\alpha$. Then it's easy to see that 
\beq
W_\gamma B_{\gamma'} = \cdots \sigma^\alpha_x \sigma^\alpha_z \cdots 
= -  \cdots \sigma^\alpha_z \sigma^\alpha_x \cdots = - B_{\gamma'} W_\gamma.
\eeq
More generally, it follows by an identical calculation that
\beq
W_\gamma B_{\gamma'} = (-)^{\#(\gamma\cap\gamma')}B_{\gamma'} W_\gamma.
\eeq
And, by what we said before (that the local Hamiltonians are in fact particular line operators for closed curves), it immediately follows that 
\beq
W_\gamma h_b = \begin{cases} 
- h_b W_\gamma, & \text{$\gamma$ ends at~$b$} \\
+ h_b W_\gamma, & \text{otherwise}.
\end{cases}
\eeq
In light of these relations, a clear picture emerges: Line operators supported on closed curves are products of local Hamiltonians, and so commute with~$H$ and all the $h$'s. Line operators supported on open curves, on the other hand, commute with all Hamiltonians of the same color, and all those of the opposite color \emph{except} those two on the endpoints of the curve. At those two squares, the line operator flips the sign of the eigenvalue of~$h$. It follows that, if (for instance) $W_\gamma$ acts on the ground state, a new energy eigenstate is produced in which the energy is bumped up by 2.

Energy eigenstates in the toric code, therefore, can be represented by imagining quasiparticles of two species living on the white and black squares. Each particle number is conserved modulo 2, so that we can imagine pairs of particles connected by line operators $W$ and $B$ on open curves. Furthermore, the particles can be moved around freely; the physical state is independent of how the line operators are drawn, and we may deform them however we like. 

However, while the physical state does not depend on how the lines are drawn, its \emph{phase} may! And this phase is not insignificant; rather, since one can imagine processes that act on the state by the corresponding phase, it is possible in principle to measure such relative phases---and they can have important consequences.  
For example, in the usual spin/statistics theorem, it is assumed that the physical state corresponding to a collection of indistinguishable particles is invariant under exchange of particles. However, this does not preclude the possibility that the phase of the wavefunction transforms according to a character of~$S_n$. Since the character needs to be defined for all~$n$, the only meaningful possibilities are the trivial or sign characters; these correspond to bosonic and fermionic particles, and have far-reaching physical consequences, including the Pauli exclusion principle.

In two spatial dimensions, though, $S_n$ is replaced by the braid group $B_n$, and more interesting possibilities---\emph{anyonic exchange statistics}---are possible. Indeed, we've constructed the phase already in the above: Imagine that there's a white and a black quasiparticle in a region of the lattice, connected by strings to distant partners. It is clear that, if the white quasiparticle is transported once around the black one, the intersection number of the strings changes by exactly one, so that the state acquires a phase $-1$!

So we've shown that excitations of the toric code model correspond to anyonic quasiparticles, appearing at the end of the line operators $W$ and~$B$. 
But furthermore---and this is really the punch line---when the spatial topology is nontrivial, so that there are closed curves that do not bound, the corresponding line operators commute with all local Hamiltonians. But they form an algebra that \emph{cannot be represented nontrivially} in any energy eigenspace! If periodic boundary conditions are imposed on the lattice, and $X$ and~$Y$ are curves (of either type $\gamma$ or~$\gamma'$) representing a basis of the first homology, then the relevant algebra is generated by $W_{X,Y}$ and~$B_{X,Y}$, subject to the relations
\beq
\begin{gathered}
W_{X,Y}^2 = B_{X,Y}^2 = 1, \\
W_X B_Y = - B_Y W_X, \quad
W_Y B_X = - B_X W_Y, \\
[W_{X},B_{X}] = [W_Y, B_Y] = 0.
\end{gathered}
\eeq
And this algebra, by a simple analogue of the Stone--von Neumann theorem, admits an essentially unique four-dimensional irreducible representation. (The trivial representation is not admissible, because of the presence of the nontrivial exchange phase! To put it differently, we've already demanded that a central element in our algebra be represented by the fixed nontrivial number $-1$.) Indeed, this algebra is a simple example of the quantum torus algebra, taken over~$\Z/2\Z$, and the quantum torus is itself nothing more than the exponentiated form of the canonical commutation relations $[q,p] = i\hbar$. This family of related constructions sometimes go under the name of \emph{Heisenberg groups}~\cite[for example]{weil}.

Thus, \emph{every} eigenspace of the toric code has a fourfold degeneracy, coming from the action of the algebra of line operators wrapping nontrivial cycles. The anyonic statistics of the quasiparticles ensure that the phases in that algebra are nontrivial, and therefore that it cannot be trivially represented. 
A common statement in the physics literature is that the toric code model reproduces abelian Chern--Simons theory, with two $U(1)$ gauge fields, and matrix of coupling constants
\beq
\kappa = \begin{bmatrix} 0 & 2 \\ 2 & 0 \end{bmatrix}.
\label{eq:Kmat}
\eeq
The content of this statement is that precisely the same algebra of line operators is present in both models, and ensures identical ground-state degeneracies (topological Hilbert space dimensions). The Chern--Simons interaction acts by attaching a magnetic flux to each electrically charged particle, so that the exchange statistics are fractionalized. With off-diagonal coupling as in~\eqref{eq:Kmat}, one obtains a phase when one species of particle (Wilson line operator) is transported around the other. 

One thus sees that we've returned to the idea that symmetries are responsible for every robust degeneracy in quantum-mechanical systems. Properly understood, a symmetry is nothing other than the action of a collection of operators commuting with the Hamiltonian. In quantum field theory, though, those operators may be local, or have extended support in spacetime~\cite{generalized}. Topological order, at least in spatial dimension two, arises when the model admits quasiparticles that admit mutually nontrivial anyonic exchange statistics. One can then consider operators defined by creating quasiparticles, transporting them around one another (or along nontrivial paths in the spatial topology) to obtain a nontrivial transformation of the state, and then annihilating them to return to the vacuum. The resulting algebras of line operators must be represented nontrivially, due to the existence of phases, and the corresponding symmetry is therefore always ``spontaneously broken:'' One obtains a robust, topology-dependent vacuum degeneracy, which is described by a three-dimensional topological quantum field theory. 
And the use of an axiomatic picture of topological quantum field theory to study the zoology of possible phases---and, in particular, to compute the groups of invertible phases that we mentioned above---is the subject of Dan's lectures in this volume. We hope we've managed to provide some small window onto why these questions were formulated in the physics literature, how a physicist might begin to think about them, and why they are of such great interest.


\section*{Acknowledgements}

First and foremost, this paper could not have been written without Max~Metlitski, on whose lecture at the Bozeman summer school it is in part based, and who was an invaluable help in producing the final version.
The author is also deeply grateful to Ryan Grady and David Ayala, for their hard work organizing such a wonderful meeting and for the invitation to prepare this paper, and to Montana State University, for its warm hospitality during the workshop.
Further thanks are due to Johannes Walcher for comments on the draft, and to Alex Turzillo for conversations.
The author's work is supported in part by the Deutsche Forschungsgemeinschaft, within the
framework of the Exzellenzinitiative an der Universit{\"a}t Heidelberg.


\bibliographystyle{amsplain}
\bibliography{spin-systems}

\end{document}